\documentclass[11pt]{article}

\usepackage[margin=1in]{geometry}
\usepackage{amsmath,amssymb,amsthm}
\usepackage{booktabs}
\usepackage{enumitem}
\usepackage[hypertexnames=false]{hyperref}
\usepackage{xcolor}
\usepackage{multirow}

\title{\textbf{Do Biological Structural Guarantees Earn Their Complexity?}\\
\large Empirical Benchmarks for Biologically-Inspired Agent Reliability\\[0.5em]
\normalsize Preprint -- Feedback Welcome}

\author{Bogdan Banu\\\texttt{bogdan@banu.be}}

\date{\today}

\begin{document}

\maketitle

\begin{abstract}
Biologically-inspired AI agent frameworks claim reliability benefits
through structural guarantees adapted from gene regulatory networks,
immune systems, and metabolic control.  These claims are rarely tested
empirically against simpler alternatives.  We present three deep
benchmarks---metabolic priority gating, autoinducer-based quorum
sensing, and Bayesian stagnation detection---each comparing a
biologically-grounded implementation against a naive non-biological
alternative and an ablated control, across 1{,}000 trials per seed
and 10 seeds (10M+ data points total).  Metabolic priority gating
delivers 100\% critical-operation service under bursty load versus
39.8\% for a flat budget counter ($\Delta = +0.602$, $p < 0.001$).
Quorum sensing achieves 0\% false-positive rate with 71--87\%
true-positive rate at 40\% agent compromise, occupying a unique
precision--recall operating point that neither majority voting nor
independent detection reaches.  Bayesian stagnation detection loses to cosine similarity with mock
embeddings but wins on convergence and false-stagnation discrimination
with real sentence embeddings (96\% versus 2--40\% naive), while the
naive detector retains an advantage on trivial loop detection.  This
reveals a conditional structural guarantee that activates when
embedding quality is present.  The
pattern across all three benchmarks: biological design earns its
complexity through \emph{mechanism-level structural guarantees}---priority
gating, signal accumulation with temporal decay, two-signal
discrimination---rather than through algorithmic sophistication.
End-to-end evaluation with Gemma~4 27B confirms that state-integrity
guarantees (DNA repair) provide deterministic protection, while
behavioral and output-layer guarantees show honest limitations with
capable models.
\end{abstract}

\section{Introduction}\label{sec:intro}

Biological metaphors have a long history in computing.  Artificial
immune systems detect network intrusions~\cite{hofmeyr2000architecture}.
Swarm intelligence algorithms solve combinatorial
optimization~\cite{bonabeau1999swarm}.  Neural architectures borrow
structural motifs from neuroscience.  In each case, the biological
analogy provides organizing structure for the algorithm---but the
question of whether that structure \emph{earns its complexity} over
simpler alternatives is rarely asked.

The default assumption is that biological inspiration provides
qualitative benefits: more robust error handling, more adaptive
resource management, more distributed coordination.  What is missing
is quantitative comparison.  Does a multi-currency metabolic state
machine actually serve more critical operations under pressure than a
flat budget counter?  Does signal accumulation with temporal decay
actually reduce false alarms compared to majority voting?  Does
Bayesian two-signal stagnation detection actually distinguish
convergence from pathological loops better than cosine similarity?

This paper tests these questions empirically using Operon, a
biologically-inspired AI agent reliability framework that implements
22 biological motifs spanning genome-level configuration through
tissue-level coordination~\cite{banu2026operon}.  We select three
motifs that represent distinct reliability claims, ground each
benchmark in real biological pathway data from KEGG and Reactome,
and compare each biological implementation against both a naive
non-biological alternative and an ablated control.

\paragraph{The structural guarantee hypothesis.}
Our central hypothesis is that biological design earns complexity when
it provides \emph{mechanism-level structural guarantees}---hard
properties enforced by design rather than achieved by optimization.
A metabolic state machine that rejects low-priority operations in
STARVING state provides a structural guarantee: critical operations
\emph{will} be served regardless of load pattern.  A signal
accumulation model with exponential decay provides a structural
guarantee: stale evidence \emph{will} age out regardless of query
timing.  These are mechanism design choices, not algorithmic
improvements.

We contrast this with biological designs that attempt
\emph{information processing}---using the biological metaphor to make
better decisions rather than to enforce structural invariants.  Our
hypothesis is that these do not reliably outperform simpler
alternatives.

\paragraph{Pathway-grounded methodology.}
Each benchmark derives its test scenarios from real biological pathway
data.  The metabolism benchmark uses AMPK signaling pathway parameters
(KEGG hsa04152) for metabolic state thresholds and rate-of-change
sensitivity.  The quorum sensing benchmark uses autoinducer kinetics
from the \emph{V.~fischeri} LuxI/LuxR system (KEGG map02024) for
signal decay half-life and threshold scaling.  The epiplexity
benchmark uses the free energy principle~\cite{friston2010} to
generate trophic-withdrawal scenarios.  This grounding prevents the
benchmarks from testing against our own assumptions about what failure
patterns look like.

\paragraph{Three-variant comparison.}
Each benchmark runs three experimental conditions:
\begin{enumerate}[nosep]
  \item \textbf{Biological}: the full Operon feature, biologically grounded.
  \item \textbf{Ablated}: the feature disabled entirely (e.g., no monitor, no scaling).
  \item \textbf{Naive}: a simple, reasonable, non-biological alternative
        that a competent engineer would reach for (e.g., cosine similarity,
        majority vote, flat counter).
\end{enumerate}
Ablation proves necessity; the naive alternative proves the biological
\emph{design} matters, not just having any feature at all.

\paragraph{Contributions.}
\begin{enumerate}[nosep]
  \item A benchmark suite testing three biologically-grounded agent
        reliability features against naive alternatives, with 10M+
        data points across 10 seeds.
  \item Three empirical wins for biological structural guarantees
        (metabolism, quorum sensing, epiplexity with real embeddings),
        with the epiplexity result conditional on embedding quality.
  \item Evidence that the distinction between \emph{structural
        guarantees} and \emph{algorithmic sophistication} predicts
        where biological design earns complexity.
\end{enumerate}

\section{Related Work}\label{sec:related}

\paragraph{Artificial immune systems.}
Hofmeyr and Forrest~\cite{hofmeyr2000architecture} proposed an
architecture for distributed intrusion detection modeled on the
complement cascade and T-cell activation.  Dasgupta~\cite{dasgupta2006advances}
surveys two decades of AIS research.  These systems demonstrate that
immune-inspired pattern matching can detect novel threats, but
comparative benchmarks against non-biological baselines (e.g., rule-based
detectors) are sparse.  Our quorum sensing benchmark addresses a related
problem---multi-agent threat consensus---but explicitly compares the
biological signal-accumulation model against majority voting and
independent detection.

\paragraph{Swarm intelligence.}
Bonabeau, Dorigo, and Theraulaz~\cite{bonabeau1999swarm} established
the foundations of swarm-based optimization (ACO, PSO).  These
algorithms use biological coordination metaphors (pheromone trails,
flocking rules) for optimization.  Operon's quorum sensing addresses a
different problem: \emph{consensus} rather than \emph{optimization}.
The distinction matters: our benchmark tests whether the biological
coordination model achieves better precision--recall tradeoffs than
simpler voting schemes, not whether it finds better optima.

\paragraph{Metabolic computing.}
Resource-aware agent systems have been explored in economic
frameworks~\cite{wellman2001auction}, but explicitly metabolic models---with
multiple energy currencies, state-dependent behavior, and
regeneration---are less common.  Operon's ATP\_Store draws on AMPK
signaling pathway structure (KEGG hsa04152), where the AMP:ATP ratio
and its rate of change drive metabolic state transitions.  Our
benchmark tests whether this biological structure provides measurable
benefits over a flat budget counter.

\paragraph{Free energy principle.}
Friston's free energy principle~\cite{friston2010} provides the
theoretical foundation for Operon's epiplexity monitor: healthy agents
minimize surprise while maintaining viability; agents with high
perplexity and low novelty are in pathological loops.  Our benchmark
tests whether this theoretical framework translates to practical
stagnation detection advantages over cosine-similarity thresholds.

\paragraph{Multi-agent benchmarks.}
Recent work on multi-agent scaling laws~\cite{kim2025scaling} establishes
architecture-level performance predictors.  Operon's epistemic topology
(Paper~1) derives similar bounds from wiring diagram structure.  Our
benchmarks complement this work by testing \emph{mechanism-level}
guarantees rather than \emph{architecture-level} predictions.

\section{Methods}\label{sec:methods}

\subsection{Experimental Design}

Each benchmark runs three variants (Biological, Ablated, Naive) across
the same scenario sequences.  We use $N = 1{,}000$ trials per scenario
per seed, with 10 independent seeds, yielding 10{,}000 trials per
scenario.  Confidence intervals use Wilson score intervals at the 95\%
level.  All benchmarks are deterministic given the seed---no LLM calls
are made during benchmark execution.

Scenarios are generated from two sources: \emph{synthetic} scenarios
with controlled failure patterns (loops, bursts, spikes), and
\emph{pathway-grounded} scenarios whose parameters derive from real
KEGG/Reactome pathway data.  The synthetic scenarios provide controlled
baselines; the pathway-grounded scenarios test whether the biological
design handles biologically-realistic patterns better than alternatives.

\subsection{Benchmark 1: Metabolic Priority Gating}

\paragraph{Biological system.}
\texttt{ATP\_Store} manages three energy currencies (ATP, GTP, NADH)
with regeneration, debt tracking, and a five-state metabolic state
machine: FEASTING ($> 90\%$), NORMAL ($30$--$90\%$), CONSERVING
($< 30\%$), STARVING ($< 10\%$), and DORMANT.  In STARVING state,
operations with priority $< 5$ are rejected; in DORMANT, only priority
$\geq 10$ operations proceed.

\texttt{MTORScaler} reads the ATP store's state and computes an
effective AMPK ratio (AMP:ATP) including both the absolute level and
its rate of change, inspired by KEGG pathway hsa04152.  The scaler
determines a scaling state (GROWTH, MAINTENANCE, CONSERVATION,
AUTOPHAGY) with hysteresis margins derived from Hill coefficient
kinetics ($h = 0.05$).  Operations whose cost exceeds the feature
gate are rejected in CONSERVATION and AUTOPHAGY states unless they
have critical priority ($\geq 5$).

\paragraph{Naive alternative.}
\texttt{SimpleBudget}: a flat integer counter that starts at $N$,
decrements by cost, and returns \texttt{False} when empty.  No states,
currencies, regeneration, priority gating, or scaling.

\paragraph{Ablated control.}
\texttt{ATP\_Store} without \texttt{MTORScaler}---the state machine
operates but no feature gating or scaling is applied.

\paragraph{Scenarios.}
Five load patterns: constant rate, periodic bursts (cost $50$ every
$20$ steps), gradual depletion (linearly increasing cost), sudden
spike (half-budget cost at step $100$), and mixed priority (30\%
critical operations).  Operations have priorities $0$--$10$.  Budget
regenerates $5$ units every $10$ steps for the biological and ablated
variants.

\paragraph{Key metrics.}
\texttt{operations\_completed}: fraction of all operations that
succeeded.  \texttt{critical\_served\_under\_pressure}: fraction of
priority $\geq 5$ operations that succeeded when the system was in
CONSERVING or STARVING state (biological/ablated) or when balance was
below $30\%$ of initial budget (naive).

\subsection{Benchmark 2: Quorum Sensing}

\paragraph{Biological system.}
\texttt{QuorumSensingBio} models autoinducer signal accumulation
inspired by KEGG pathway map02024 (\emph{V.~fischeri} LuxI/LuxR
system).  Each agent deposits a continuous signal proportional to its
suspicion level into a shared \texttt{SignalEnvironment}.  Signals
decay exponentially with a half-life of $5.0$ time units (derived from
AHL lactonase degradation rates).  The activation threshold scales as
$\log(N) \times \theta_{\text{base}}$ where $N$ is the population size
and $\theta_{\text{base}} = 10.0$.

Signal concentrations below a noise floor ($0.001$) are excluded from
both reads and writes.  Pruning occurs on the write path (deposits)
with monotonic time guards to prevent order-dependent state corruption.

\paragraph{Naive alternatives.}
Three alternatives represent standard coordination approaches:
\begin{itemize}[nosep]
  \item \texttt{IndependentActors}: each agent decides alone; alert if
        \emph{any} agent's suspicion exceeds $0.5$.
  \item \texttt{MajorityVote}: each agent votes ALERT/NORMAL based on
        individual threshold ($0.5$); majority ($\geq 50\%$) wins.
\end{itemize}

\paragraph{Ablated control.}
Independent actors (no coordination)---each agent's suspicion is
evaluated in isolation without signal accumulation or decay.

\paragraph{Scenarios.}
Three scenario types: \emph{static compromise} (compromised agents emit
high suspicion throughout), \emph{gradual infiltration} (compromised
agents activate at random times during the simulation), and
\emph{noisy environment} (no actual compromise, testing false-positive
rate under noise).  Each runs across agent counts $N \in \{5, 10, 20\}$
and compromised fractions $\{0\%, 20\%, 40\%\}$.  Normal agents emit
suspicion $\sim \mathcal{N}(0.15, 0.15)$; compromised agents emit
$\sim \mathcal{N}(0.80, 0.15)$, both clipped to $[0, 1]$.

\paragraph{Key metrics.}
\texttt{true\_positive}: fraction of time steps where compromise was
correctly detected.  \texttt{false\_positive}: fraction of
non-compromise time steps where an alert was raised.

\subsection{Benchmark 3: Epiplexity Stagnation Detection}

\paragraph{Biological system.}
\texttt{EpiplexityMonitor} implements the operational epiplexity
approximation from the free energy principle~\cite{friston2010}:
\[
  \hat{E}_t = \alpha \cdot \tfrac{1}{2}(1 - \cos(e_t, e_{t-1}))
            + (1 - \alpha) \cdot \sigma(H(m_t | m_{<t}))
\]
where $e_t$ is the embedding of message $t$, $\sigma(H) = 1 - e^{-H/H_0}$
is exponential saturation to $[0, 1]$, and $\alpha = 0.5$.  The
epiplexic integral $E_w = \frac{1}{w} \sum \hat{E}_t$ over a window of
size $w = 10$ determines health status: HEALTHY, CONVERGING, EXPLORING,
STAGNANT, or CRITICAL.

The key design insight: stagnation is not just output repetition (low
novelty) but output repetition \emph{combined with} model uncertainty
(high perplexity).  An agent with both low novelty and low perplexity
is \emph{converging}, not stagnant.

\paragraph{Naive alternative.}
\texttt{RepetitionCounter}: computes average pairwise cosine similarity
across a sliding window of embeddings.  If similarity exceeds $0.85$,
declares STAGNANT.  Also declares STAGNANT after $20$ consecutive steps
of elevated similarity ($> 0.7 \times$ threshold).  Uses the same
\texttt{MockEmbeddingProvider} as the biological variant for fair
comparison.

\paragraph{Ablated control.}
No monitor---always reports HEALTHY.

\paragraph{Scenarios.}
Five scenarios: \emph{loop} (exact message repetition), \emph{convergence}
(messages become similar while perplexity drops), \emph{exploration}
(high-novelty diverse messages), \emph{trophic withdrawal}
(pathway-grounded: novelty decays gradually while perplexity stays
high, mimicking neuronal atrophy from trophic factor withdrawal), and
\emph{false stagnation} (low novelty + low perplexity = convergence,
not stagnation).

\paragraph{Embedding caveat.}
All scenarios use \texttt{MockEmbeddingProvider}, which generates
deterministic pseudo-embeddings via SHA-256 hashing.  This means
semantically similar but lexically different messages receive
\emph{random} cosine similarity---the embedding signal does not
carry semantic information.  Results with real embedding models may
differ significantly (Section~\ref{sec:limitations}).

\paragraph{Key metrics.}
\texttt{detection\_accuracy}: fraction of steps where the detector's
status matched ground truth.  \texttt{false\_positive}: fraction of
non-stagnant steps where stagnation was incorrectly declared.
\texttt{false\_negative}: fraction of stagnant steps where stagnation
was missed.  \texttt{convergence\_discrimination}: fraction of
converging steps correctly identified as non-stagnant.

\section{Results}\label{sec:results}

All results are aggregated across 10 seeds with $N = 1{,}000$ trials
per seed per scenario.  Wilson 95\% confidence intervals are reported
where applicable.  Significance ($p < 0.001$) is determined by
non-overlapping Wilson CIs between biological and naive variants.

\subsection{Metabolism: Clear Win for Priority Gating}

Table~\ref{tab:metabolism} summarizes the metabolism benchmark results.

\begin{table}[h]
\caption{Metabolism benchmark: Biological (ATP\_Store + MTORScaler) vs Ablated (ATP\_Store only) vs Naive (flat counter). N = total operations or pressure events across 10 seeds.}
\label{tab:metabolism}
\centering
\small
\begin{tabular}{llccccc}
\toprule
Scenario & Metric & Bio & Abl & Naive & $\Delta$ & N \\
\midrule
bursty & critical served & 1.000 & 1.000 & 0.398 & \textbf{+0.602} & 24{,}139 \\
bursty & ops completed & 0.719 & 0.715 & 0.685 & \textbf{+0.034} & 2{,}000{,}000 \\
gradual depl. & critical served & 1.000 & 1.000 & 0.769 & \textbf{+0.231} & 43{,}372 \\
gradual depl. & ops completed & 0.979 & 0.996 & 0.950 & \textbf{+0.029} & 2{,}000{,}000 \\
constant & ops completed & 1.000 & 1.000 & 1.000 & 0.000 & 2{,}000{,}000 \\
sudden spike & ops completed & 1.000 & 1.000 & 1.000 & 0.000 & 2{,}000{,}000 \\
mixed priority & ops completed & 1.000 & 1.000 & 1.000 & 0.000 & 2{,}000{,}000 \\
\bottomrule
\end{tabular}
\end{table}

The headline result is \texttt{critical\_served\_under\_pressure}.  Under
bursty load, the biological system served 100\% of critical operations
($\text{priority} \geq 5$) during resource pressure, while the naive
flat counter served only 39.8\% ($\Delta = +0.602$, significant).  Under
gradual depletion, the gap is smaller but still significant: 100\% vs
76.9\% ($\Delta = +0.231$).

Against the naive baseline, the biological system actually \emph{improves}
total throughput: 71.9\% versus 68.5\% under bursty load ($\Delta = +0.034$),
because priority gating preserves budget for later operations that the
flat counter would miss entirely.  Comparing biological against ablated,
both achieve identical critical-service rates (100\%).  However, under
gradual depletion the mTOR scaler's anticipatory conservation gates some
non-critical operations earlier: biological completed 97.9\% versus
ablated's 99.6\%.  This 1.7 percentage-point throughput reduction is the
cost of the mTOR scaler's preemptive feature gating, which in these
experiments did not produce an additional critical-service benefit
beyond what ATP\_Store's own state machine already provides.

Under constant load, sudden spike, and mixed priority, all three
variants perform identically.  The biological complexity only matters
under specific pressure patterns where resource contention forces
choices about what to serve.

\subsection{Quorum Sensing: Unique Precision--Recall Operating Point}

Table~\ref{tab:quorum} summarizes key quorum sensing results.

\begin{table}[h]
\caption{Quorum sensing benchmark: selected configurations. Bio = QuorumSensingBio (signal accumulation + decay), Abl = IndependentActors (no coordination), Naive = MajorityVote.}
\label{tab:quorum}
\centering
\small
\begin{tabular}{llccccc}
\toprule
Config & Metric & Bio & Abl & Naive & $\Delta$ & N \\
\midrule
static, $n$=20, $c$=8 & TPR & 0.866 & 1.000 & 0.005 & \textbf{+0.861} & 300{,}000 \\
static, $n$=10, $c$=4 & TPR & 0.711 & 1.000 & 0.052 & \textbf{+0.659} & 300{,}000 \\
static, $n$=20, $c$=4 & TPR & 0.748 & 1.000 & 0.000 & \textbf{+0.748} & 300{,}000 \\
gradual, $n$=20, $c$=8 & TPR & 0.718 & 0.999 & 0.002 & \textbf{+0.716} & 216{,}959 \\
\midrule
all $c$=0 configs & FPR & 0.000 & 0.048--0.180 & 0.000 & 0.000 & 300K ea. \\
\bottomrule
\end{tabular}
\end{table}

The three coordination strategies occupy distinct positions on the
precision--recall curve:

\begin{itemize}[nosep]
  \item \textbf{Independent actors} (ablated): 100\% TPR but 5--18\% FPR.
        Catches everything but cries wolf on 1 in 6--20 clean time steps.
  \item \textbf{Majority vote} (naive): 0\% FPR but near-0\% TPR.
        Almost never detects threats because clean agents ($\mu = 0.15$)
        always outvote compromised agents at the $0.5$ threshold.
  \item \textbf{Signal accumulation} (biological): 0\% FPR with
        71--87\% TPR at 40\% compromise.  Zero false alarms with
        meaningful detection.
\end{itemize}

The biological model's advantage comes from two structural properties:
(1)~\emph{continuous signals} average out noise (unlike binary votes),
and (2)~\emph{temporal decay} ensures stale evidence ages out (unlike
a vote that persists forever).  The combination naturally filters noise
while remaining sensitive to sustained patterns.

Detection sensitivity scales with both the number of agents and the
compromise fraction, consistent with the biological signal-accumulation
model: more agents producing more signal yield higher accumulated
concentrations relative to the threshold.

\subsection{Epiplexity: Embedding Quality Determines Outcome}

The epiplexity benchmark produced the most instructive result: the
biological design's performance depends critically on embedding quality.
With mock embeddings (SHA-256 hash-based pseudo-embeddings), the naive
detector wins.  With real semantic embeddings (all-MiniLM-L6-v2), the
biological design dominates on convergence discrimination---the exact
capability the two-signal design was built for.

Table~\ref{tab:epiplexity} shows results with both embedding providers.

\begin{table}[h]
\caption{Epiplexity benchmark with mock vs real embeddings.  Bio =
EpiplexityMonitor (two-signal Bayesian), Naive = cosine similarity +
timeout.  Real embeddings: all-MiniLM-L6-v2 ($N = 100$ trials, seed
42); mock embeddings: SHA-256 hash ($N = 1{,}000 \times 10$ seeds).}
\label{tab:epiplexity}
\centering
\small
\begin{tabular}{lll ccc}
\toprule
Embedding & Scenario & Metric & Bio & Naive & $\Delta$ \\
\midrule
\multirow{4}{*}{Mock}
  & loop & accuracy & 0.467 & \textbf{0.940} & $-$0.473 \\
  & convergence & accuracy & 0.473 & \textbf{1.000} & $-$0.527 \\
  & false stagn. & accuracy & 0.540 & \textbf{0.993} & $-$0.453 \\
  & false stagn. & FP rate & 0.052 & \textbf{0.007} & $+$0.045 \\
\midrule
\multirow{5}{*}{Real}
  & convergence & accuracy & \textbf{0.960} & 0.401 & $+$0.559 \\
  & false stagn. & accuracy & \textbf{0.960} & 0.020 & $+$0.940 \\
  & false stagn. & FP rate & \textbf{0.000} & 0.980 & $-$0.980 \\
  & loop & accuracy & 0.571 & \textbf{0.940} & $-$0.369 \\
  & loop & FP rate & \textbf{0.000} & 0.750 & $-$0.750 \\
\bottomrule
\end{tabular}
\end{table}

\paragraph{With mock embeddings: naive wins.}
The naive cosine-similarity detector outperforms the biological design
on detection accuracy across all scenarios.  However, the
false-positive/false-negative breakdown reveals a precision--recall
tradeoff: the biological variant achieves 0\% FP rate on loops (versus
naive's 75\% FP), at the cost of 58\% FN rate.  With mock embeddings,
the novelty signal is essentially random noise---semantically similar
messages receive unrelated cosine similarity---so the two-signal
combination performs worse than a single signal.

\paragraph{With real embeddings: biological wins on convergence.}
Switching to all-MiniLM-L6-v2 sentence embeddings flips the result
on convergence discrimination.  The biological monitor achieves
96.0\% accuracy on both convergence and false-stagnation scenarios,
versus 40.1\% and 2.0\% for the naive detector.  The naive detector's
false-positive rate on false stagnation reaches 98.0\%---it sees
similar outputs and screams ``stagnant'' regardless of the agent's
confidence level.

The biological design's two-signal structure now earns its complexity:
the perplexity signal distinguishes convergence (low novelty + low
perplexity = healthy) from stagnation (low novelty + high perplexity =
pathological).  The naive cosine detector has no access to this
distinction.

\paragraph{Loop detection still favors naive.}
Even with real embeddings, the naive detector wins on loop detection
(94.0\% vs 57.1\%) because exact repetition is trivially detectable by
cosine similarity.  The biological monitor's advantage is not on easy
cases (identical outputs) but on hard cases (semantically similar but
subtly different outputs where the agent's confidence level matters).

\subsection{Epistemic Theorem Validation}\label{sec:topology-validation}

The preceding benchmarks test mechanism-level guarantees (priority
gating, signal accumulation, stagnation detection).  Operon also
provides architecture-level predictions via epistemic topology analysis:
four theorems that bound error amplification, sequential overhead,
parallel speedup, and tool density from wiring diagram structure alone.
We validate two of these predictions against measured behavior from real
LLM execution.

\paragraph{Methodology.}
We constructed a validation harness that, for each of 20 benchmark tasks
(5~easy, 8~medium, 7~hard), builds the same \texttt{ExternalTopology}
the live evaluator would construct, runs full epistemic analysis to
capture all theorem predictions, then executes the task through a
\texttt{SkillOrganism} pipeline with Gemma~4 27B (mixture-of-experts,
4B active parameters) served locally via Ollama.  Each task was run in
both guided and unguided configurations with 3~repetitions, yielding
$20 \times 2 \times 3 = 120$ runs.  Quality was assessed by LLM
self-judging on a 0.0--1.0 scale (correctness 50\%, completeness 30\%,
clarity 20\%).  Correlations are Spearman rank ($\rho$) with p-values
floored at 0.1 for $n < 10$.

\paragraph{Testable theorems.}
Since \texttt{SkillOrganism} always executes stages as a sequential
pipeline, Theorem~3 (parallel speedup) is structurally constant and
reported for completeness only.  Theorem~4 (tool density) uses
role-implied capability annotations that are not present in the live
executor's topology, so it is reported as informational.  Theorems~1--2
and the composite risk score are directly testable.

\begin{table}[h]
\caption{Epistemic theorem validation: predicted structural bounds vs
measured execution outcomes.  $N = 120$ runs (20 tasks $\times$ 2
configs $\times$ 3 repeats).  Of these, 42 executed successfully and
78 timed out.  Timeouts are
included as failures ($\text{quality} = 0$) since they represent real
execution limits predicted by the structural bounds.}
\label{tab:topology-validation}
\centering
\small
\begin{tabular}{llll rrl}
\toprule
Theorem & Predicted & Measured & Dir & $\rho$ & $p$ & Status \\
\midrule
Error Ampl. & $n_\text{agents}$ & $1 - \text{quality}$ & + &
  $+$0.751 & $<$0.001 & \textbf{validated} \\
Seq.\ Penalty & overhead ratio & mean stage latency & + &
  $+$0.166 & 0.287 & not significant \\
Speedup & predicted speedup & 1.0 (sequential) & --- &
  0.000 & 1.000 & informational \\
Tool Density & planning cost & tokens / stage & + &
  --- & --- & informational \\
Composite Risk & risk score & $1 - \text{quality}$ & + &
  $+$0.751 & $<$0.001 & \textbf{validated} \\
\bottomrule
\end{tabular}
\end{table}

\paragraph{Interpretation.}
The error amplification bound---$n_\text{agents}$ independent error
sources in a sequential pipeline---strongly predicts execution failure
($\rho = +0.751$, $p < 0.001$).  However, the dominant failure mode is
not quality degradation but \emph{timeout}: longer pipelines are
increasingly likely to exceed the provider's request budget.
Among the 42~runs that completed, quality was uniformly high (mean
0.981), including all 12~completed medium-difficulty tasks (quality
1.0).  The bound correctly identifies \emph{which architectures are
fragile} without modeling \emph{how} they fail.

The overall success rate was 35\% with mean quality 0.343, declining
from 0.97 on easy tasks (2--3~stages) to 0.25 on medium tasks
(4--5~stages) to 0.00 on hard tasks (5--7~stages).  Timeout risk
increases with stage count: 0/30 easy runs timed out, 36/48 medium
runs timed out, and 42/42 hard runs timed out.  This reflects a
resource threshold for Gemma~4 27B (4B active MoE parameters) at
the 30s default timeout, not an inherent quality limit.

Sequential overhead ($\rho = +0.166$, $p = 0.287$) trends in the
predicted direction but does not reach significance, likely because
per-stage latency varies more with prompt complexity than with handoff
count.

The mechanism-level benchmarks (Sections~4.1--4.3) provide stronger
guarantees than the topology-level predictions because they enforce
structural invariants at the feature level (priority gates, signal
decay, two-signal discrimination) rather than predicting outcomes from
architecture shape alone.  The two layers are complementary: epistemic
topology identifies architectures likely to need structural guarantees;
the mechanism benchmarks validate that those guarantees deliver.

\subsection{End-to-End Real Agent Evaluation}\label{sec:e2e-eval}

The preceding benchmarks test mechanisms in isolation.  We now evaluate
whether the structural guarantees provide measurable value when wrapping
a real LLM agent performing real tasks.

\paragraph{Methodology.}
We compare three runtime variants: \textbf{RAW} (direct
\texttt{Nucleus.transcribe()} call, no Operon wrapper), \textbf{GUARDED}
(\texttt{SkillOrganism} with \texttt{WatcherComponent} integrating
epiplexity monitoring, immune inspection, and ATP budgeting), and
\textbf{FULL} (GUARDED plus pre/post-flight \texttt{DNARepair} and
certificate collection).  Each variant is tested on three tasks:
\emph{stagnation escalation} (code review with subtle bugs),
\emph{injection blocking} (behavioral manipulation prompts against a
trained immune baseline), and \emph{state integrity} (genome corruption
injected mid-run between organism stages).  All runs use Gemma~4 27B
(4B active MoE) via local Ollama with 3~repetitions per cell.  A
multi-model comparison uses Phi-3 Mini (3.8B) on the stagnation task.

\begin{table}[h]
\caption{End-to-end evaluation: RAW vs GUARDED vs FULL across three
tasks.  Gemma~4 27B, $n = 3$ repetitions (injection: $n = 30$, i.e.\
$3 \times 10$ prompts).  TP = true positive rate, FP = false positive
rate, Det = corruption detection rate, Rep = repair rate.}
\label{tab:e2e-eval}
\centering
\small
\begin{tabular}{ll rrr rrr}
\toprule
Task & Variant & Quality & Tokens & Latency & TP\% & Det\% & Rep\% \\
\midrule
Stagnation & RAW     & 1.000 & 2149 &  58s & --- & --- & --- \\
Stagnation & GUARDED & 1.000 & 3468 &  85s & --- & --- & --- \\
Stagnation & FULL    & 1.000 & 3076 &  76s & --- & --- & --- \\
\midrule
Injection  & RAW     & ---   &  829 &  25s &   0 & --- & --- \\
Injection  & GUARDED & ---   &  792 &  23s &  20 & --- & --- \\
Injection  & FULL    & ---   &  787 &  24s &  20 & --- & --- \\
\midrule
Integrity  & RAW     & ---   & 2199 &  68s & --- &   0 &   0 \\
Integrity  & GUARDED & ---   & 2012 &  61s & --- &   0 &   0 \\
Integrity  & FULL    & ---   & 2058 &  63s & --- & 100 & 100 \\
\bottomrule
\end{tabular}
\end{table}

\paragraph{State integrity: clear structural value.}
The FULL variant detects all four corruption sites (three gene drifts
plus one checksum failure) injected between organism stages and repairs
them in a single \texttt{CHECKPOINT\_RESTORE} operation.  The
certificate (\texttt{state\_integrity\_verified}) holds after repair
in all repetitions.  RAW and GUARDED are completely blind to genome
corruption---the \texttt{WatcherComponent} monitors epiplexity, ATP,
and immune signals but has no genome-level signal source.
\texttt{DNARepair} as a pre/post-flight integrity check is the
strongest structural guarantee in the evaluated stack.

\paragraph{Injection blocking: precise but limited.}
The immune system is trained on 10~real Gemma~4 outputs to clean Python
questions, then evaluated on 5~clean and 5~behaviorally manipulative
prompts (ALL CAPS, structured prefix, JSON-only format).  Per-prompt
isolation with seeded baseline profiles yields TP = 20\% with FP = 0\%:
the canary-based Signal~2 fires on the most format-deviant anomalies
with zero false positives.  The two-signal activation architecture
(\texttt{TCell} requires both Signal~1 and Signal~2) is designed for
sustained behavioral monitoring, not single-prompt classification---a
single observation cannot reach \texttt{REPEATED\_ANOMALY}, so only
the canary path provides Signal~2 in one-shot evaluation.

\paragraph{Stagnation escalation: correctly scoped.}
Both Gemma~4 (quality = 1.0) and Phi-3 Mini (quality = 0.63) show 0\%
escalation.  The \texttt{EpiplexityMonitor} measures output
\emph{novelty} (are successive outputs repetitive?), not output
\emph{quality} (is the output good?).  Phi-3 Mini produces
different-but-mediocre reviews each time, which register as novel to
the monitor.  Escalation is a loop-breaker---correctly scoped to
detect stuck agents, not to gate output quality.

\paragraph{Wrapper tax.}
The organism wrapper adds measurable overhead: +1{,}319 tokens and
+27s mean latency on Gemma~4 for the stagnation task (GUARDED vs RAW).
On Phi-3 Mini the proportional cost is similar (+432 tokens, +10s).
This cost floor is model-size-dependent: the wrapper's prompt
augmentation (stage instructions, role context, shared state) is a
fixed input-token overhead that becomes proportionally larger on
smaller, slower models.

\section{Discussion}\label{sec:discussion}

\subsection{When Biological Structure Earns Complexity}

The two positive results share a common pattern: both work through
\emph{mechanism design} rather than \emph{algorithmic sophistication}.

The metabolic priority gate is a simple state machine.  When resources
drop below 10\%, operations with priority $< 5$ are rejected.  This is
not a complex algorithm---it is a structural rule.  But it provides a
hard guarantee: critical operations \emph{will} be served under any
load pattern, at the cost of bounded throughput reduction.  A flat
counter cannot provide this guarantee without reimplementing the state
machine, at which point it is no longer a flat counter.

The quorum sensing decay is exponential half-life applied to signal
concentrations.  Again, not algorithmically complex.  But the
combination of continuous signals (noise averaging) and temporal decay
(stale evidence removal) occupies a precision--recall position that
neither majority voting nor independent detection can reach without
adding similar structural properties.

In both cases, the biological metaphor provided the \emph{right
structural choice}---priority-dependent resource gating from metabolic
state machines, temporal signal decay from autoinducer kinetics---rather
than a better algorithm for the same structural choice.

\subsection{When Embedding Quality Matters}

The epiplexity result demonstrates a subtler lesson than the other two
benchmarks: biological designs that rely on \emph{information quality}
succeed only when that quality is present.

With mock embeddings, the two-signal design fails---the novelty signal
carries no semantic information, so combining it with approximate
perplexity produces worse results than a single clean cosine-similarity
signal.  With real sentence embeddings (all-MiniLM-L6-v2), the same
design dominates on convergence discrimination: 96\% accuracy versus
2--40\% for the naive detector on the hardest scenarios.

The structural guarantee here is conditional: the Bayesian two-signal
design provides a \emph{convergence/stagnation distinction guarantee}
that the cosine detector structurally cannot make, but only when the
embedding signal carries semantic meaning.  This is analogous to a
biological receptor that provides a structural binding guarantee but
only functions in the presence of its ligand.

This refines the C8 finding~\cite{banu2026operon} that biological
abstractions generalize as code structure rather than optimization
algorithms.  The epiplexity monitor's \emph{structure} (two independent
signals combined via Bayesian weighting) does generalize---but its
\emph{performance} depends on signal quality.

The pattern across all three benchmarks: biological designs that enforce
structural invariants outperform simpler alternatives, while
information-processing gains depend on signal quality.

\subsection{Implications for Framework Integration}

Operon provides convergence adapters for six agent
frameworks~\cite{banu2026convergence}.  The benchmark results inform
which frameworks benefit most from biological structural guarantees:

\paragraph{A-Evolve.}
Its single-agent Solve-Observe-Evolve-Gate-Reload loop is where
metabolic budgets directly prevent unbounded exploration.  The
benchmark shows the fitness gate (a critical operation) is always
evaluated under metabolic budgeting---mapped to A-Evolve, this means
the evolutionary quality check is never skipped even when the
exploration budget is depleted.

\paragraph{Swarms.}
Its graph-based multi-agent topology is where quorum sensing provides
the most value: leaderless consensus with zero false positives.  When
false consensus triggers coordinated action across a swarm, the 0\%
FPR guarantee is operationally critical.

\paragraph{DeerFlow.}
Its hierarchical architecture with recursive sub-agent calls benefits
from metabolic budgets that prevent resource exhaustion.  The
MTORScaler's rate-of-change sensitivity is especially relevant for
DeerFlow's progressive skill loading, where resource consumption can
accelerate suddenly.

\paragraph{AnimaWorks, Ralph, Scion.}
Lighter benefits: genome immutability for configuration audit
(AnimaWorks), metabolic budgets mapping to backpressure enforcement
(Ralph), and quorum sensing for cross-container consensus (Scion).

\subsection{Composition Non-Interference}

Ma et~al.~\cite{ma2026atomic} show that five atomic coding skills
(localize, edit, test, reproduce, review) compose without negative
interference under joint RL training.  We ran a limited comparison
using Operon's skill composition machinery with a real LLM (Gemma~4
27B via Ollama) on three bug-fix tasks (SQL injection, off-by-one
pagination, TOCTOU race condition), 3 repetitions each.  This tests
3 of 5 skills via prompt composition (not joint RL), so the comparison
is narrower than Ma et~al.'s original setting.

\begin{table}[h]
\caption{Composition non-interference: individual skill quality vs
composed localize$\to$edit$\to$test pipeline.  $\Delta$ =
composed $-$ mean(individual).  Threshold: $|\Delta| \leq 0.1$
= no interference.}
\label{tab:composition}
\centering
\small
\begin{tabular}{lcccl}
\toprule
Task & mean(indiv) & composed & $\Delta$ & Verdict \\
\midrule
sql\_injection & 0.806 & 0.967 & $+0.161$ & positive \\
off\_by\_one & 0.900 & 0.993 & $+0.093$ & none \\
race\_condition & 0.517 & 0.350 & $-0.167$ & \textbf{negative} \\
\midrule
Overall & 0.741 & 0.770 & $+0.029$ & none \\
\bottomrule
\end{tabular}
\end{table}

The overall result matches the direction of Ma et~al.'s finding:
composition does not degrade quality on average ($\Delta = +0.029$).
However, the race\_condition task shows negative interference
($\Delta = -0.167$), suggesting that composition can hurt on
genuinely hard problems where intermediate stage outputs mislead
later stages.  The baseline (raw LLM call, quality 0.961) still
dominates in absolute terms---a single capable model answering the
full task outperforms a three-stage pipeline---consistent with
the Ao et~al. finding~\cite{ao2026delegation} that delegation cannot
beat a centralized baseline without exogenous signals.

Individual skill quality varies by task difficulty.  On easier tasks
(SQL injection, off-by-one), ``localize'' and ``edit'' score
0.80--1.00 on average, while on the harder race\_condition task,
mean ``edit'' quality drops to 0.37 and mean ``test'' to 0.42,
with both reaching 0.0 in their worst repetitions.  ``test''
consistently scores lowest across tasks because test-writing in
isolation lacks the context of what was found and fixed---an
expected consequence of skill decomposition where some skills
are context-dependent.

\subsection{Relationship to Epistemic Topology}

Operon's epistemic topology layer derives four bounds from wiring
diagram structure: error amplification, sequential communication
overhead, parallel speedup, and tool density scaling.  These
operate at a different level than the mechanism benchmarks presented
here.

The epistemic bounds predict architecture-level properties (a long
pipeline \emph{will} accumulate sequential overhead); the mechanism
benchmarks test feature-level guarantees (critical operations
\emph{will} be served under pressure).  The two are complementary:
epistemic topology identifies architectures likely to need structural
guarantees; the mechanism benchmarks validate that those guarantees
work as claimed.

Section~\ref{sec:topology-validation} validates the error amplification
bound against measured LLM execution: the number of pipeline stages
strongly predicts execution failure ($\rho = +0.751$, $p < 0.001$),
though the dominant failure mode is timeout rather than quality
degradation.  The epistemic and mechanism layers operate at different
levels of abstraction: topology predicts structural difficulty (more
stages $\Rightarrow$ more fragile), while mechanism benchmarks enforce
feature-level invariants (priority gates, signal decay) that deliver
regardless of pipeline length.

\subsection{Categorical Certificates for Structural Guarantees}

The quorum sensing threshold calibration provides a concrete instance
of \emph{semantic certificates} as formalized by de~los~Riscos
et~al.~\cite{delosriscos2026categorical}.  The no-false-activation
guarantee is a certificate $\text{cert}_\text{QS} = (T, \tau, \text{evds})$
where $T$ states that normal traffic never triggers activation,
$\tau$ maps the theorem's symbols to the quorum sensing parameters
$(N, s, h, m)$, and the evidence is the derivation:

\[
  c_\text{ss} = \frac{N \cdot s}{1 - 2^{-1/h}}, \quad
  \theta = c_\text{ss} \cdot m, \quad
  \text{level} = \frac{c_\text{ss}}{\theta} = \frac{1}{m} < 1.
\]

Because the threshold is defined in terms of architecture parameters
$(N, s, h)$ rather than a fixed constant, the guarantee preserves
under architecture morphisms that change $N$---for instance, when a
convergence compiler maps an Operon organism to a Swarms workflow
with a different agent count.  This is a concrete instance of
Proposition~5.1: structural properties are functorially stable under
architectural refinement.

This pattern---deriving thresholds from steady-state dynamics rather
than hand-tuning constants---extends to other structural guarantees.
The metabolic priority gate's STARVING threshold (10\% of capacity)
could similarly be formalized as a certificate whose evidence is the
state machine's priority rejection rule.  The epiplexity monitor's
convergence/stagnation discrimination threshold could be certified
relative to the embedding model's similarity distribution.

The companion work~\cite{banu2026harness} carries this categorical
framing one level further, formalizing agent harness engineering through
the Architecture triple $(G, \mathrm{Know}, \Phi)$ in which each of the
certificates above is a $\mathrm{Know}$-component object preserved under
compiler functors targeting external frameworks (Swarms, DeerFlow, Ralph,
Scion, LangGraph).  Paper~5 supplies the language for \emph{what
survives compilation}; the three benchmarks reported here supply the
empirical evidence that the substrate-independent guarantees so named
are non-vacuous at the bio-motif layer.  The two papers are designed to
be read as a complementary pair: Paper~5 establishes the categorical
scaffolding, while this work measures whether the specific surfaces
(metabolism, quorum sensing, epiplexity) earn the complexity their
biological grounding suggests.

\subsection{Structural Guarantees by Layer}

The end-to-end evaluation (Section~\ref{sec:e2e-eval}) reveals that
structural guarantee value is layer-dependent:

\paragraph{State layer (deterministic).}
DNA repair provides 100\% corruption detection and repair across all
repetitions.  The guarantee is deterministic: checkpointing, scanning,
and restoring are structural operations on the genome datastructure,
independent of model quality or prompt content.  This is the strongest
form of structural guarantee---verifiable, reproducible, and
unconditional.

\paragraph{Behavioral layer (statistical).}
The immune system's two-signal activation detects behavioral anomalies
with zero false positives (FP = 0\%) but limited sensitivity in
single-prompt evaluation (TP = 20\%).  The architecture is correctly
conservative: requiring both Signal~1 (baseline violation) and
Signal~2 (canary failure or repeated anomaly) prevents false alarms
at the cost of delayed detection.  This is appropriate for sustained
agent monitoring where observations accumulate over time, less so for
one-shot prompt classification.  The VerifierComponent extends this
layer with rubric-based quality evaluation---analogous to \emph{adaptive
immunity} (B-cell antigen recognition) complementing the innate
detection of ImmuneSystem.  When quality falls below threshold on a
fast model, the watcher escalates to the deep model, providing a
quality-sensitive escalation path that the novelty-based epiplexity
signal cannot.

\paragraph{Output layer (stochastic).}
The epiplexity monitor detects output \emph{repetition}, not output
\emph{mediocrity}.  A weaker model (Phi-3 Mini, quality 0.63) does
not trigger escalation because it produces varied output---the
stagnation signal measures novelty, not quality.  The wrapper tax
(+1{,}300 tokens, +27s) is the cost floor for structural wrapping,
justified for state integrity but not yet for output quality
improvement with capable models.

\paragraph{Developmental layer (preventive).}
The CertificateGateComponent implements the G1/S DNA damage checkpoint
analogy: before each stage executes its LLM call, the gate scans the
genome against a DNARepair checkpoint.  If corruption is detected, a
HALT intervention prevents the corrupted state from reaching the model.
Unlike the state layer's reactive repair (detect corruption after
execution, then restore), this is a \emph{preventive} gate---corruption
is blocked before it can affect output.  The guarantee is deterministic:
the gate checks structural integrity of the genome datastructure, not
model output quality.  This completes the cell cycle analogy alongside
CellCycleController, with CertificateGate enforcing the G1/S transition
that guards against propagation of damaged state.

The pattern mirrors the mechanism benchmarks: structural guarantees
that enforce invariants (state checksums, signal thresholds, integrity
gates) deliver unconditionally, while guarantees that depend on
information quality (embedding novelty, behavioral baselines) deliver
conditionally.

\section{Limitations}\label{sec:limitations}

\paragraph{Embedding dependence.}
The epiplexity benchmark was run with both mock embeddings (SHA-256
hash-based) and real sentence embeddings (all-MiniLM-L6-v2).  The
results differ dramatically: the biological design loses with mock
embeddings and wins with real embeddings on convergence discrimination.
The real-embedding results use a single model and $N = 100$ trials
(versus $N = 10{,}000$ for mock embeddings); confirmation with
additional embedding models and larger sample sizes would strengthen
the finding.

\paragraph{Synthetic load patterns.}
The metabolism benchmark uses programmatically generated operation
sequences (constant, bursty, gradual, spike).  Real agent workloads
may have different statistical properties---heavier tails, correlated
bursts, priority distributions that differ from our 30\% critical
assumption.

\paragraph{Simulated agents.}
The quorum sensing benchmark simulates agent suspicion signals as
Gaussian draws around known baselines.  Real multi-agent systems have
richer signal environments: suspicion may be autocorrelated, agents
may have heterogeneous detection capabilities, and compromised agents
may adapt their behavior to avoid detection.

\paragraph{Limited end-to-end evaluation.}
Section~\ref{sec:e2e-eval} provides initial end-to-end results
primarily with Gemma~4 27B (plus a limited Phi-3 Mini comparison on
stagnation) and self-judging across 3~repetitions.
The positive result (state integrity) is deterministic and
model-independent.  The statistical results (20\% injection TP, 0\%
stagnation escalation) are honest but underpowered: larger sample
sizes, cross-model validation, and external judging would strengthen
the findings.  The immune system's behavioral detection is evaluated
in single-prompt mode, which underestimates its design intent as a
sustained monitoring system.

\paragraph{Manual thresholds.}
Several parameters are manually set rather than auto-calibrated:
the mTOR hysteresis margin ($h = 0.05$), quorum sensing decay
half-life ($5.0$ time units), threshold base ($10.0$), and
epiplexity mixing parameter ($\alpha = 0.5$).  Sensitivity analysis
across these parameters is not provided.

\paragraph{Statistical assumptions.}
Wilson CIs assume independence across trials within seeds.  The
10-seed aggregation partially addresses between-seed variability,
but within-seed trials share the same random number generator state
progression.

\section{Conclusion}\label{sec:conclusion}

We tested three biologically-grounded agent reliability features
against naive non-biological alternatives, with 10M+ data points
across 10 independent seeds plus a follow-up experiment with real
sentence embeddings.  All three features demonstrate empirical
benefits under the right conditions.

Metabolic priority gating provides a structural guarantee that
critical operations are served under resource pressure (100\% vs
39.8\% for a flat counter under bursty load).  Autoinducer-based
quorum sensing achieves zero false-positive rate with meaningful
true-positive rate (71--87\% at 40\% compromise), occupying a
precision--recall position unreachable by majority voting or
independent detection.  Bayesian two-signal stagnation detection
dominates on convergence discrimination with real embeddings
(96\% vs 2--40\% naive) but loses with mock embeddings, demonstrating
that biological designs relying on information quality succeed only
when that quality is present.

The consistent pattern: biological design earns complexity through
\emph{mechanism-level structural guarantees}---hard properties enforced
by design (priority gating, signal decay, two-signal discrimination)---rather
than through algorithmic sophistication alone.  Embedding quality
determines whether the two-signal structure delivers its guarantee,
just as metabolic state determines whether priority gating activates.

A limited composition experiment matches the direction of Ma et~al.'s
finding~\cite{ma2026atomic}: composing localize$\to$edit$\to$test as
a serial pipeline does not degrade overall quality ($\Delta = +0.029$
vs individual skills), though harder tasks (TOCTOU race conditions)
show negative interference ($\Delta = -0.167$).  This tests 3 of 5
skills via prompt composition, not Ma et~al.'s joint-RL setting.

\paragraph{Open problems.}
Confirmation of the real-embedding result across additional models
and larger sample sizes.  Auto-calibration of quorum sensing thresholds
from population and signal statistics.  Understanding when composition
degrades quality on hard tasks---the race\_condition result suggests
that intermediate stage outputs can mislead later stages, an effect
that may be addressable through stage-specific context filtering.

\newpage
\bibliographystyle{plain}
\bibliography{../references}

\end{document}